\documentclass{Interspeech}
\usepackage{cite,colortbl,tabularx}
\hypersetup{colorlinks=true,citecolor=green,linkcolor=red,urlcolor=magenta}
\newcommand{\first}[1]{\textbf{#1}}
\newcommand{\second}[1]{\textbf{\textit{#1}}}
\newcommand{\customparagraph}[1]{\noindent\textbf{#1}}
\newcolumntype{C}{>{\centering\arraybackslash}X}
\makeatletter
\newcommand\footnoteref[1]{\protected@xdef\@thefnmark{\ref{#1}}\@footnotemark}
\makeatother
\interspeechcameraready

\title{FasterVoiceGrad: Faster One-step Diffusion-Based Voice Conversion\\
  with Adversarial Diffusion Conversion Distillation}

\author{Takuhiro}{Kaneko}
\author{Hirokazu}{Kameoka}
\author{Kou}{Tanaka}
\author{Yuto}{Kondo}

\affiliation[nocounter]{}{NTT, Inc.}{Japan}
\email{takuhiro.kaneko@ntt.com}

\begin{document}

\maketitle

\begin{abstract}
  A diffusion-based voice conversion (VC) model (e.g., VoiceGrad) can achieve high speech quality and speaker similarity; however, its conversion process is slow owing to iterative sampling. FastVoiceGrad overcomes this limitation by distilling VoiceGrad into a one-step diffusion model. However, it still requires a computationally intensive content encoder to disentangle the speaker's identity and content, which slows conversion. Therefore, we propose \textit{FasterVoiceGrad}, a novel one-step diffusion-based VC model obtained by simultaneously distilling a diffusion model and content encoder using \textit{adversarial diffusion conversion distillation (ADCD)}, where distillation is performed in the conversion process while leveraging adversarial and score distillation training. Experimental evaluations of one-shot VC demonstrated that \textit{FasterVoiceGrad} achieves competitive VC performance compared to FastVoiceGrad, with 6.6--6.9 and 1.8 times faster speed on a GPU and CPU, respectively.\footnote{\label{foot:samples}Audio samples are available at \url{https://www.kecl.ntt.co.jp/people/kaneko.takuhiro/projects/fastervoicegrad/}.}
\end{abstract}

\section{Introduction}
\label{sec:introduction}

Voice conversion (VC) is a technique that converts one voice to another while preserving the linguistic content.
VC has been extensively studied owing to its wide range of applications.
Early studies began with parallel VC, which trains a converter using a parallel corpus.
However, parallel corpora are often impractical to collect.
Non-parallel VC, which learns a converter without a parallel corpus, has been proposed to overcome this drawback.
However, non-parallel VC has difficulty in learning owing to the absence of direct supervision; nonetheless, deep generative model-based VC, such as (variational) autoencoder (VAE/AE)~\cite{DKingmaICLR2014}-based VC~\cite{CHsuAPSIPA2016,HKameokaTASLP2019,KTanakaSSW2023,KQianICML2019,JChouIS2019,YHChenICASSP2021,DWangIS2021}, generative adversarial network (GAN)~\cite{IGoodfellowNIPS2014}-based VC~\cite{TKanekoEUSIPCO2018,HKameokaSLT2018,TKanekoIS2019,HKameokaTASLP2020c,TKanekoICASSP2021,YLiIS2021}, flow~\cite{LDinhICLRW2015}-based VC~\cite{JSerraNeurIPS2019}, and diffusion~\cite{JSohlICML2015}-based VC~\cite{HKameokaTASLP2024,VPopovICLR2022,HYChoiIS2023,HYChoiAAAI2024,JHaiIS2024,TKanekoIS2024}, have mitigated this difficulty.

This study focused on \textit{diffusion-based non-parallel VC} owing to its ease of data collection and strong performance in terms of speech quality and speaker similarity.
Despite these advantages, a diffusion model has the drawback of slow conversion due to iterative sampling (e.g., tens of steps).
This property is disadvantageous compared with those of other VC methods, such as VAE-based and GAN-based VC, because they can complete the conversion in a one-step feedforward process.

\begin{figure}[t]  
  \centering
  \includegraphics[width=0.99\linewidth]{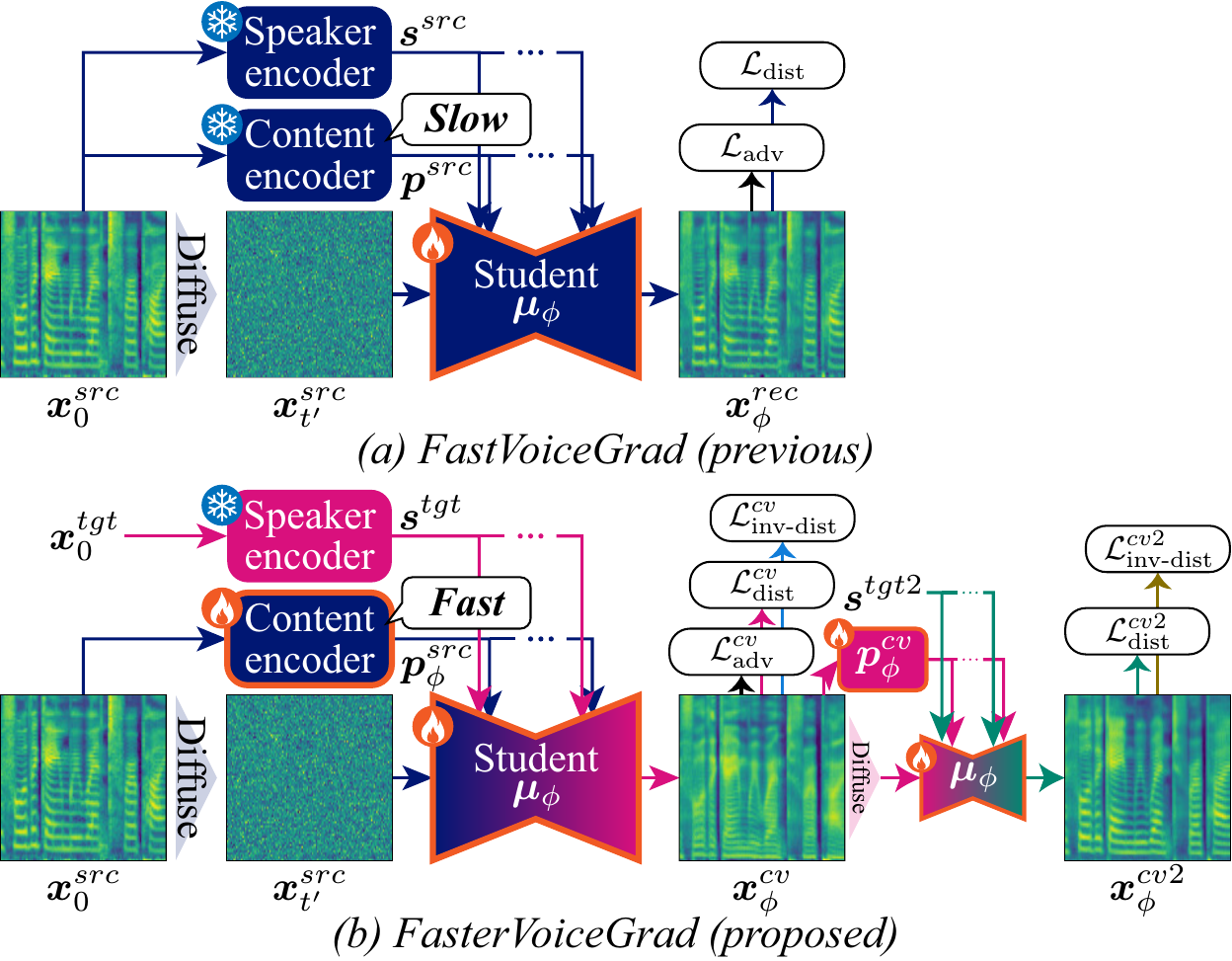}
  \vspace{-3mm}
  \caption{Comparison of training processes between (a) FastVoiceGrad (previous) and (b) \textit{FasterVoiceGrad} (proposed).
    (a) FastVoiceGrad distills the main reverse diffusion module $\bm{\mu}_{\phi}$ in the \textbf{reconstruction} process.
    (b) In contrast, \textit{FasterVoiceGrad} distills not only the main reverse diffusion module $\bm{\mu}_{\phi}$ but also the content encoder $\bm{p}_{\phi}$ in the \textbf{conversion} process.
    This change enables learning of conversion while preventing learning an identity mapping through $\bm{p}_{\phi}$.
    The details of each notation and process are explained in Sections~\ref{sec:preliminary} and \ref{sec:proposal}.}
  \label{fig:teaser}
  \vspace{-5.5mm}
\end{figure}

FastVoiceGrad~\cite{TKanekoIS2024} was proposed to overcome this drawback.
This is a one-step diffusion-based VC model distilled from the multi-step VoiceGrad model~\cite{HKameokaTASLP2024} and achieves a VC performance comparable to that of VoiceGrad while reducing the number of sampling steps from tens to one by leveraging adversarial training~\cite{IGoodfellowNIPS2014} and score distillation training~\cite{ASauerECCV2024}.
However, its computational cost reduction effect is limited to the main reverse diffusion process, and the computational cost of the content encoder, which is essential for disentangling content and speaker identity, remains high.\footnote{The computational cost of a content encoder is not negligible because a computationally intensive model, such as a conformer~\cite{AGulatiIS2020}-based bottleneck feature extractor (BNE)~\cite{SLiuTASPL2021}, HuBERT~\cite{WNHsuTASLP2021}, or XLS-R~\cite{ABabuIS2022}, is commonly used.}

To overcome this limitation, we consider simultaneously distilling not only the main reverse diffusion process but also the content encoder.
A straightforward solution is to replace the original content encoder with a faster trainable module and train it using the FastVoiceGrad method.
However, as shown in Figure~\ref{fig:teaser}(a), FastVoiceGrad training is performed in the \textit{reconstruction} process; therefore, the training can end up learning an \textit{identity mapping}, that is, simply outputting the input through a trainable content encoder path.\footnote{Note that reconstruction is not trivial for the main reverse diffusion process because its input is not clean but diffused data.}
This problem can be mitigated using direct distillation, which aligns the output of the replaced content encoder with that of the original.
However, as discussed in Section~\ref{subsec:model_analysis}, this method results in a performance decrease when a less computationally intensive model is used.

Alternatively, we propose \textit{FasterVoiceGrad}, a novel one-step diffusion-based VC model obtained by simultaneously distilling the main reverse diffusion process and content encoder using \textit{adversarial diffusion conversion distillation (ADCD)}, where distillation is conducted in the \textit{conversion} process, \textit{not} in the \textit{reconstruction} process, as shown in Figure~\ref{fig:teaser}(b).
This change enables the learning of conversion while preventing the learning of a simple identity mapping.
Furthermore, we introduce \textit{reconversion score distillation} for content preservation and \textit{inverse score distillation} for speaker emphasis.

\textit{FasterVoiceGrad} was experimentally validated in one-shot any-to-any VC.
Model analysis demonstrated the importance of each component and the superiority of \textit{FasterVoiceGrad} over direct distillation.
A comparison with FastVoiceGrad showed that \textit{FasterVoiceGrad} achieves competitive performance compared to FastVoiceGrad while speeding up the conversion.

The rest of this paper is organized as follows:
Section~\ref{sec:preliminary} reviews VoiceGrad and FastVoiceGrad, upon which \textit{FasterVoiceGrad} was built.
Section~\ref{sec:proposal} explains \textit{FasterVoiceGrad}.
Section~\ref{sec:experiments} presents experimental results.
Finally, Section~\ref{sec:conclusion} concludes the paper and provides remarks on future research.

\section{Preliminary: VoiceGrad/FastVoiceGrad}
\label{sec:preliminary}

\subsection{VoiceGrad}
\label{subsec:voicegrad}

VoiceGrad~\cite{HKameokaTASLP2024} is a denoising diffusion probabilistic model (DDPM)~\cite{JHoNeurIPS2020}-based non-parallel VC model.
It uses a mel-spectrogram $\bm{x}$ as the conversion target and converts it based on speaker embedding $\bm{s}$ (e.g.,~\cite{YJiaNeurIPS2018}) and content embedding $\bm{p}$ (e.g.,~\cite{SLiuTASPL2021}).
In the conversion process, $\bm{x}$ and $\bm{p}$ are extracted from the source speaker's speech, while $\bm{s}$ is extracted from the target speaker's speech to convert the speaker characteristics.
Hereafter, where necessary, we use superscripts $src$ and $tgt$ to denote the source and target speakers, respectively.
In VoiceGrad, the diffusion process, reverse diffusion process, and training objective are defined as follows:

\smallskip
\customparagraph{Diffusion process.}
In this process, the input mel-spectrogram $\bm{x}_0$ ($= \bm{x}$) is transformed into noise $\bm{x}_T \sim \mathcal{N}(\bm{0}, \bm{I})$ over $T$ steps ($T = 1000$ in practice).
Using the reproductivity of the normal distribution and a reparameterization trick~\cite{DKingmaICLR2014}, the $t$-step diffused data $\bm{x}_t$ ($t \in \{ 1, \dots, T \}$) can be written as
\begin{flalign}
  \label{eq:diffusion}
  \bm{x}_t = \sqrt{\bar{\alpha}_t} \bm{x}_0 + \sqrt{1 - \bar{\alpha}_t} \bm{\epsilon},
\end{flalign}
where $\bar{\alpha}_t = \prod_{i=1}^t \alpha_i$, $1 - \alpha_i$ denotes the noise variance in the $i$-th step, and $\bm{\epsilon} \sim \mathcal{N}(\bm{0}, \bm{I})$.

\smallskip
\customparagraph{Reverse diffusion process.}
In this process, $\bm{x}_T$ is gradually denoised towards $\bm{x}_0$.
The denoising process for each reverse diffusion step is written as
\begin{flalign}
  \label{eq:reverse_diffusion}
  \bm{\mu}_{\theta}(\bm{x}_t, t, \bm{s}, \bm{p}) = \frac{1}{\sqrt{\alpha_t}} \left( \bm{x}_t - \frac{1 - \alpha_t}{\sqrt{ 1 - \bar{\alpha}_t }} \bm{\epsilon}_{\theta}(\bm{x}_t, t, \bm{s}, \bm{p}) \right),
\end{flalign}
where $\bm{\epsilon}_{\theta}$ is a denoising function with parameter $\theta$.
This process is performed based on $\bm{s}^{src}$ and $\bm{p}^{src}$ during training (i.e., \textit{reconstruction} is conducted), while it is performed based on $\bm{s}^{tgt}$ and $\bm{p}^{src}$ during \textit{conversion}.

\smallskip
\customparagraph{Training objective.}
$\bm{\epsilon}_{\theta}$ is optimized to \textit{reconstruct} the noise introduced during the diffusion process using the following objective:
\begin{flalign}
  \label{eq:ddpm_loss}
  \mathcal{L}_{\textnormal{DDPM}}(\theta) = \mathbb{E}_{\bm{x}_0^{src}, t, \bm{\epsilon}} [ \lVert \bm{\epsilon} - \bm{\epsilon}_{\theta}(\bm{x}_t^{src}, t, \bm{s}^{src}, \bm{p}^{src}) \rVert_1 ].
\end{flalign}

\subsection{FastVoiceGrad}
\label{subsec:fastvoicegrad}

FastVoiceGrad~\cite{TKanekoIS2024} distills VoiceGrad into a one-step diffusion-based VC model by using adversarial loss~\cite{IGoodfellowNIPS2014,XMaoICCV2017} and score distillation loss~\cite{ASauerECCV2024}.
Hereafter, for clarity, we use $\theta$ and $\phi$ to denote the teacher and student model parameters, respectively.
As shown in Figure~\ref{fig:teaser}(a), in FastVoiceGrad, distillation is performed during the \textit{reconstruction} process.

\smallskip
\customparagraph{Adversarial loss.}
This loss (in particular, the least-squares GAN form~\cite{XMaoICCV2017}) is written as
\begin{flalign}
  \label{eq:adv_loss_D}
  \hspace{-2mm}
  \mathcal{L}_{\textnormal{adv}}(\psi) & = \mathbb{E}_{\bm{x}_0^{src}} [ (\mathcal{D}(\mathcal{V}(\bm{x}_0^{src})) - 1)^2 + (\mathcal{D}(\mathcal{V}(\bm{x}_{\phi}^{rec})))^2 ],
  \\
  \label{eq:adv_loss_G}
  \hspace{-2mm}
  \mathcal{L}_{\textnormal{adv}}(\phi) & = \mathbb{E}_{\bm{x}_0^{src}} [ (\mathcal{D}(\mathcal{V}(\bm{x}_{\phi}^{rec})) - 1)^2 ],
\end{flalign}
where $\mathcal{D}$ is a discriminator with parameter $\psi$, $\mathcal{V}$ is a vocoder, and $\bm{x}_{\phi}^{rec} = \bm{\mu}_{\phi}(\bm{x}_{t'}^{src}, t', \bm{s}^{src}, \bm{p}^{src})$ is the result of one-step reverse diffusion using the student model $\bm{\mu}_{\phi}$.
Here, $\bm{x}_{t'}^{src}$ is the $t'$-step diffused $\bm{x}_0^{src}$ ($t' = 950$ in practice) and $\bm{\mu}_{\phi}$ is calculated in the same manner as in Equation~\ref{eq:reverse_diffusion}.
In this formulation, adversarial training is performed in the time domain to obtain $\bm{x}_{\phi}^{rec}$, from which a real waveform can be synthesized.
To stabilize the GAN training, the feature matching loss $\mathcal{L}_{\textnormal{FM}}(\phi)$~\cite{KKumarNeurIPS2019,JKongNeurIPS2020} (see Equation~12 in~\cite{TKanekoIS2024}) is also applied between $\bm{x}_{\phi}^{rec}$ and $\bm{x}_0$.

\smallskip
\customparagraph{Score distillation loss.}
This loss~\cite{ASauerECCV2024} is defined as
\begin{flalign}
  \label{eq:dist_loss}
  \mathcal{L}_{\textnormal{dist}}(\phi) = \mathbb{E}_{\bm{x}_0^{src}, t} [ \sqrt{\bar{\alpha}_t} \lVert \bm{x}_{\phi}^{rec} - \bm{x}_{\theta}^{rec} \rVert_1 ],
\end{flalign}
where $\bm{x}_{\theta}^{rec} = \bm{\mu}_{\theta} (\textnormal{sg}(\bm{x}_{\phi, t}^{rec}), t, \bm{s}^{src}, \bm{p}^{src})$ is obtained by diffusing and reversely diffusing $\bm{x}_{\phi}^{rec}$ using the teacher model $\bm{\mu}_{\theta}$; here, $\textnormal{sg}$ denotes a stop gradient operation, and $\bm{x}_{\phi, t}^{rec}$ is the $t$-step diffused $\bm{x}_{\phi}^{rec}$ ($t \in \{ 1, \dots, T \}$).
This loss encourages $\bm{x}_{\phi}^{rec}$ to match the data denoised by the teacher model $\bm{\mu}_{\theta}$.

\section{Proposal: FasterVoiceGrad}
\label{sec:proposal}

\textit{FasterVoiceGrad} distils both the main reverse diffusion module and content encoder to further speed up the conversion.\footnote{The speaker encoder is excluded from speedup because it runs only once before conversion and has negligible impact on conversion time.}
Specifically, in the implementation, a computationally intensive conformer~\cite{AGulatiIS2020}-based content encoder~\cite{SLiuTASPL2021}, which is frozen during the training of FastVoiceGrad, is replaced with a trainable and computationally efficient convolutional neural network (CNN) $\bm{p}_{\phi}$.
In this formulation, the \textit{reconstruction distillation} used in FastVoiceGrad is problematic because, when the content encoder is trainable, $\bm{x}$ can be easily reconstructed through the content encoder path ($\bm{x}_0^{src} \rightarrow \bm{p}_{\phi}^{src} \rightarrow \bm{x}_{\phi}^{rec}$), and the main reverse diffusion process ($\bm{x}_{t'}^{src} \rightarrow \bm{\mu}_{\phi} \rightarrow \bm{x}_{\phi}^{rec}$) can be ignored.
To avoid this issue, in \textit{FasterVoiceGrad}, we introduce \textit{adversarial diffusion conversion distillation (ADCD)}, where the distillation is performed in the \textit{conversion} process, as shown in Figure~\ref{fig:teaser}(b).
With this change, the adversarial and score distillation losses are first extended to \textit{conversion forms} (Section~\ref{subsec:extension}).
Furthermore, we introduce \textit{reconversion score distillation} for content preservation (Section~\ref{subsec:reconversion}) and \textit{inverse score distillation} for speaker emphasis (Section~\ref{subsec:inverse}).

\subsection{Extension from reconstruction to conversion}
\label{subsec:extension}

\smallskip
\customparagraph{Conversion adversarial loss.}
As shown in Figure~\ref{fig:teaser}(b), in \textit{FasterVoiceGrad}, the main reverse diffusion module $\bm{\mu}_{\phi}$ conducts \textit{conversion} instead of reconstruction and generates $\bm{x}_{\phi}^{cv} = \bm{\mu}_{\phi}(\bm{x}_{t'}, t', \bm{s}^{tgt}, \bm{p}_{\phi}^{src})$, where $\bm{s}^{tgt}$ (which is obtained by shuffling $\bm{s}^{src}$ within the same batch in practice) is used as the input speaker embedding instead of $\bm{s}^{src}$.
With this change, the adversarial losses in Equations~\ref{eq:adv_loss_D} and \ref{eq:adv_loss_G} are replaced with
\begin{flalign}
  \label{eq:cv_adv_loss_D}
  \hspace{-2mm}
  \mathcal{L}_{\textnormal{adv}}^{cv}(\psi) & = \mathbb{E}_{\bm{x}_0^{src}} [ (\mathcal{D}(\mathcal{V}(\bm{x}_0^{src})) - 1)^2 + (\mathcal{D}(\mathcal{V}(\bm{x}_{\phi}^{cv})))^2 ],
  \\
  \label{eq:cv_adv_loss_G}
  \hspace{-2mm}
  \mathcal{L}_{\textnormal{adv}}^{cv}(\phi) & = \mathbb{E}_{\bm{x}_0^{src}} [ (\mathcal{D}(\mathcal{V}(\bm{x}_{\phi}^{cv})) - 1)^2 ].
\end{flalign}
Owing to the absence of parallel data, the feature matching loss cannot be adopted between the converted and ground-truth target mel-spectrograms.
Alternatively, the feature matching loss used in FastVoiceGrad, i.e., $\mathcal{L}_{\textnormal{FM}}(\phi)$,  is used as-is.

\smallskip
\customparagraph{Conversion score distillation loss.}
Similarly, the score distillation loss in Equation~\ref{eq:dist_loss} is replaced with
\begin{flalign}
  \label{eq:cv_dist_loss}
  \mathcal{L}_{\textnormal{dist}}^{cv}(\phi) = \mathbb{E}_{\bm{x}_0^{src}, t} [ \sqrt{\bar{\alpha}_t} \lVert \bm{x}_{\phi}^{cv} - \bm{x}_{\theta}^{cv} \rVert_1 ],
\end{flalign}
where $\bm{x}_{\theta}^{cv} = \bm{\mu}_{\theta} (\textnormal{sg}(\bm{x}_{\phi, t}^{cv}), t, \bm{s}^{tgt}, \bm{p}^{src})$ is obtained by diffusing and reversely diffusing $\bm{x}_{\phi}^{cv}$ using the teacher model $\bm{\mu}_{\theta}$ with $\bm{s}^{tgt}$ and $\bm{p}^{src}$, where $\bm{x}_{\phi, t}^{cv}$ is the $t$-step diffused $\bm{x}_{\phi}^{cv}$.

\begin{figure}[t]  
  \centering
  \includegraphics[width=0.99\linewidth]{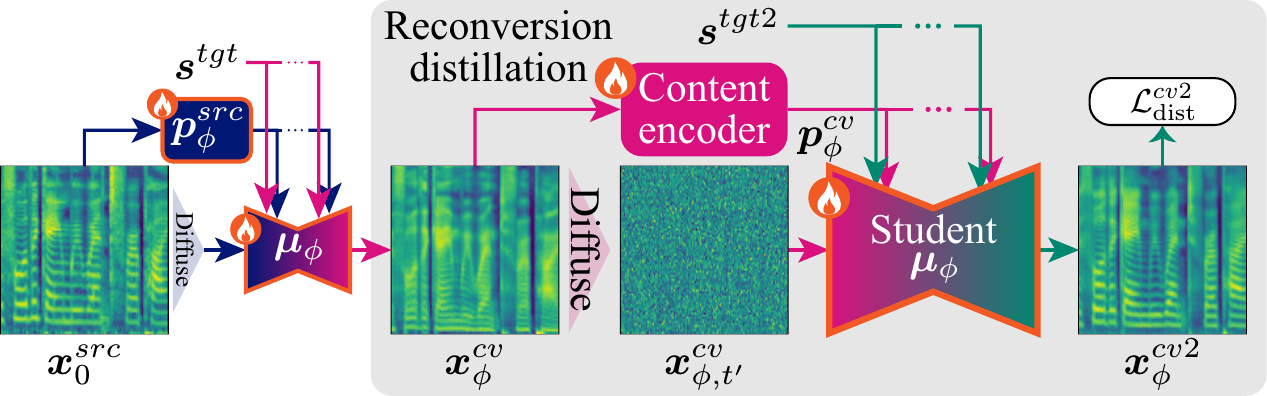}
  \vspace{-2mm}
  \caption{Process flow of reconversion score distillation.
    Reconversion score distillation loss $\mathcal{L}_{\textnormal{dist}}^{cv2}$ is applied to the twice-converted data $\bm{x}_{\phi}^{cv2}$ to enhance content preservation.}
  \label{fig:reconversion}
  \vspace{-3mm}
\end{figure}

\subsection{Content preservation by reconversion score distillation}
\label{subsec:reconversion}

As shown in Figure~\ref{fig:reconversion}, to enhance the content preservation in $\bm{x}_{\phi}^{cv}$, we conduct the conversion again and apply the score distillation loss to the twice-converted mel-spectrogram $\bm{x}_{\phi}^{cv2} = \bm{\mu}_{\phi}(\bm{x}_{\phi, t'}^{cv}, t', \bm{s}^{tgt2}, \bm{p}_{\phi}^{cv})$, where $\bm{x}_{\phi, t'}^{cv}$ is the $t'$-step diffused $\bm{x}_{\phi}^{cv}$, $\bm{s}^{tgt2}$ (which is obtained by shuffling $\bm{s}^{tgt}$ within the same batch in practice) is the speaker embedding for the second conversion, and $\bm{p}_{\phi}^{cv}$ is the content embedding extracted from $\bm{x}_{\phi}^{cv}$.
The score distillation loss for $\bm{x}_{\phi}^{cv2}$ is expressed as
\begin{flalign}
  \label{eq:rec_dist_loss}
  \mathcal{L}_{\textnormal{dist}}^{cv2}(\phi) = \mathbb{E}_{\bm{x}_0^{src}, t} [ \sqrt{\bar{\alpha}_t} \lVert \bm{x}_{\phi}^{cv2} - \bm{x}_{\theta}^{cv2} \rVert_1 ],
\end{flalign}
where $\bm{x}_{\theta}^{cv2} = \bm{\mu}_{\theta} (\textnormal{sg}(\bm{x}_{\phi, t}^{cv2}), t, \bm{s}^{tgt2}, \bm{p}^{src})$ is obtained by diffusing and reversely diffusing $\bm{x}_{\phi}^{cv2}$ using the teacher model $\bm{\mu}_{\theta}$ with $\bm{s}^{tgt2}$ and $\bm{p}^{src}$, where $\bm{x}_{\phi, t}^{cv2}$ is the $t$-step diffused $\bm{x}_{\phi}^{cv2}$.

\subsection{Speaker emphasis by inverse score distillation}
\label{subsec:inverse}

\begin{figure}[t]  
  \centering
  \includegraphics[width=0.99\linewidth]{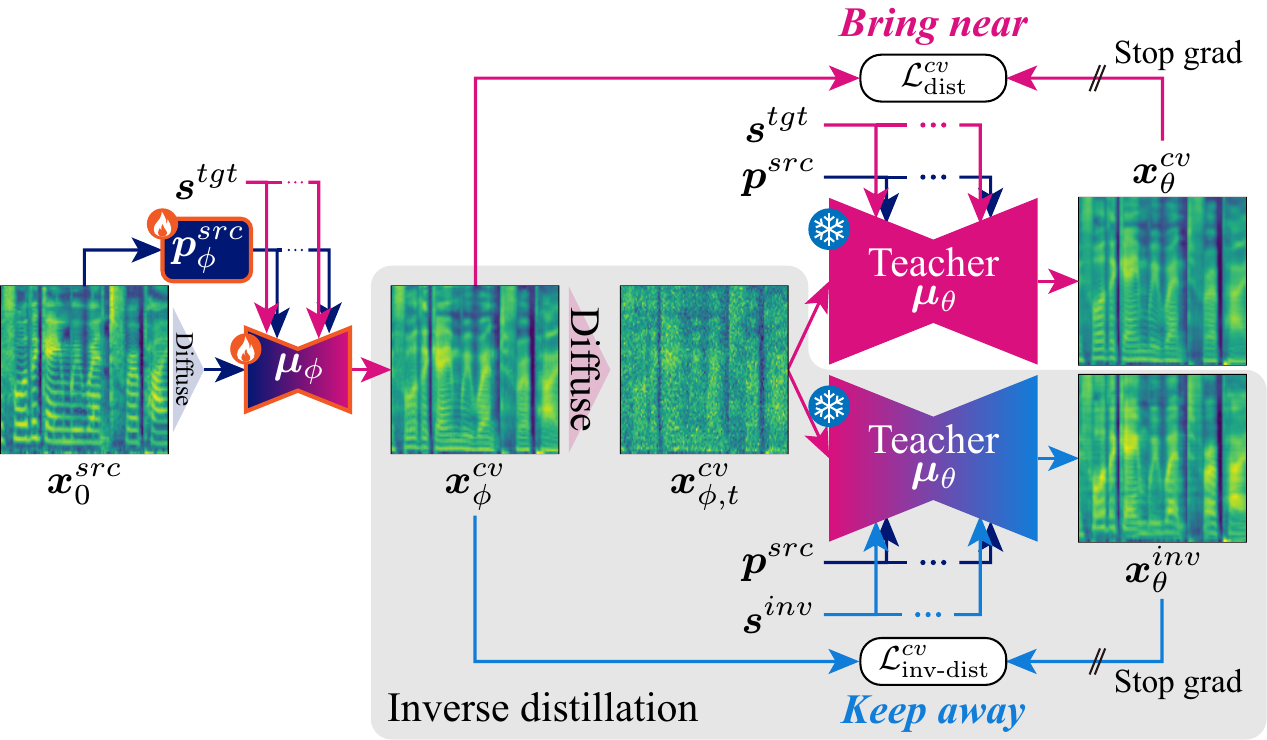}
  \vspace{-2mm}
  \caption{Process flow of inverse score distillation.
    Inverse score distillation loss $\mathcal{L}_{\textnormal{inv-dist}}^{cv}$ is applied to the converted data $\bm{x}_{\phi}^{cv}$ to enhance the speaker's identity contrastively.}
  \label{fig:inverse}
  \vspace{-3mm}
\end{figure}

As shown in Figure~\ref{fig:inverse}, to enhance the speaker's identity contrastively, we not only bring the converted mel-spectrogram $\bm{x}_{\phi}^{cv}$ closer to the target speaker's mel-spectrogram using $\mathcal{L}_{\textnormal{dist}}^{cv} (\phi)$ but also keep it away from the other speakers' mel-spectrograms using the inverse score distillation loss, which is written as
\begin{flalign}
  \label{eq:inv_dist_loss}
  \mathcal{L}_{\textnormal{inv-dist}}^{cv}(\phi) = -\mathbb{E}_{\bm{x}_0^{src}, t} [ \sqrt{\bar{\alpha}_t} \lVert \bm{x}_{\phi}^{cv} - \bm{x}_{\theta}^{inv} \rVert_1 ],
\end{flalign}
where $\bm{x}_{\theta}^{inv} = \bm{\mu}_{\theta}(\textnormal{sg}(\bm{x}_{\phi, t}^{cv}), t, \bm{s}^{inv}, \bm{p}^{src})$ is obtained by diffusing and reversely diffusing $\bm{x}_{\phi}^{cv}$ using the teacher model $\bm{\mu}_{\theta}$ with the other speaker's embedding $\bm{s}^{inv}$ (which is obtained by randomly sampling a speaker embedding other than $\bm{s}^{tgt}$ within the same batch) and $\bm{p}^{src}$.
Note that ``$-$'' is used at the top of the right-hand term because this loss aims to move the two data away from each other.
When reconversion (Section~\ref{subsec:reconversion}) is performed, a similar loss is applied to the reconverted mel-spectrogram $\bm{x}_{\phi}^{cv2}$.
We denote this loss by $\mathcal{L}_{\textnormal{inv-dist}}^{cv2}$.

\smallskip
\customparagraph{Total objective.}
The total objective is expressed as follows:
\begin{flalign}
  \label{eq:total_loss}
  \mathcal{L}_{\textnormal{ADCD}}(\phi) & = \mathcal{L}_{\textnormal{adv}}^{cv}(\phi)
                                    + \lambda_{\textnormal{FM}} \mathcal{L}_{\textnormal{FM}}(\phi)
                                    \nonumber \\
                                  & + \lambda_{\textnormal{dist}} \mathcal{L}_{\textnormal{dist}}^{cv}(\phi)
                                    + \lambda_{\textnormal{inv-dist}} \mathcal{L}_{\textnormal{inv-dist}}^{cv}(\phi)
                                    \nonumber \\
                                  & + \lambda_{\textnormal{dist}} \mathcal{L}_{\textnormal{dist}}^{cv2}(\phi)
                                    + \lambda_{\textnormal{inv-dist}} \mathcal{L}_{\textnormal{inv-dist}}^{cv2}(\phi),
  \\
  \mathcal{L}_{\textnormal{ADCD}}(\psi) & = \mathcal{L}_{\textnormal{adv}}^{cv}(\psi),
\end{flalign}
where $\mathcal{\lambda}_{\textnormal{FM}}$, $\mathcal{\lambda}_{\textnormal{dist}}$, and $\mathcal{\lambda}_{\textnormal{inv-dist}}$ are weighting hyperparameters, set to 2, 45, and 22.5, respectively, in experiments.
$\phi$ and $\psi$ are optimized by minimizing $\mathcal{L}_{\textnormal{ADCD}}(\phi)$ and $\mathcal{L}_{\textnormal{ADCD}}(\psi)$, respectively.

\section{Experiments}
\label{sec:experiments}

\subsection{Experimental setup}
\label{subsec:experimental_setup}

\customparagraph{Data.}
\textit{FasterVoiceGrad} was experimentally validated on one-shot any-to-any VC.
The main experiments (Sections~\ref{subsec:model_analysis} and \ref{subsec:comparison_with_fastvoicegrad}) were performed on the VCTK dataset~\cite{JYamagishiVCTK2019} (including utterances from 110 English speakers), and a dataset versatility analysis (Section~\ref{subsec:versatility_analysis}) was performed on the LibriTTS dataset~\cite{HZenIS2019} (including utterances from approximately 1,100 English speakers).
To evaluate unseen-to-unseen VC, ten speakers and ten sentences were excluded from the evaluation, and the remaining data were used for training.
Following previous studies~(e.g., \cite{VPopovICLR2022,TKanekoIS2024}), audio clips were downsampled to 22.05 kHz.
From these, 80-dimensional log-mel spectrograms were extracted with an FFT size of 1024, hop size of 256, and window size of 1024, and were used as the conversion targets.

\smallskip
\customparagraph{Implementation.}
The implementation of \textit{FasterVoiceGrad} follows that of FastVoiceGrad~\cite{TKanekoIS2024}, except that the content encoder is changed from a computationally intensive BNE~\cite{SLiuTASPL2021} to a trainable and efficient $\bm{p}_{\phi}$, and the objective function is modified as described in Section~\ref{sec:proposal}.
$\bm{\mu}_{\phi}$ consists of a U-Net~\cite{ORonnebergerMICCAI2015} comprising a 12-layer 1D CNN with 512 hidden channels, two downsampling/upsampling layers, gated linear unit (GLU)~\cite{YDauphinICML2017}, and weight normalization (WN)~\cite{TSalimansNIPS2016}.
$\bm{p}_{\phi}$ comprises a three-layer 1D CNN with 512 hidden channels, GLU, instance normalization~\cite{DUlyanovArXiv2016}, and WN.
$\bm{s}$ was extracted using a speaker encoder~\cite{YJiaNeurIPS2018}.
$\mathcal{V}$ is a pretrained HiFiGAN-V1 generator~\cite{JKongNeurIPS2020}.\footnote{\url{https://github.com/jik876/hifi-gan}}
$\mathcal{D}$ comprises multi-period~\cite{JKongNeurIPS2020} and multi-resolution~\cite{WJangIS2021} discriminators.
Training was performed using the Adam optimizer~\cite{DPKingmaICLR2015} with a batch size of 32, learning rate of 0.0002, $\beta_1$ of 0.5, and $\beta_2$ of 0.9 for 100 epochs on VCTK and 50 epochs on LibriTTS.

\smallskip
\customparagraph{Evaluation metrics.}
The VC performance was primarily evaluated using four objective metrics:
(1) \textit{UTMOS$\uparrow$}~\cite{TSaekiIS2022}, the predicted mean opinion score (MOS) tuned for synthesized speech,
(2) \textit{DNSMOS$\uparrow$}~\cite{CReddyICASSP2021}, the predicted MOS optimized for noise-suppressed speech,
(3) \textit{character error rate (CER)$\downarrow$} by Whisper-large-v3~\cite{ARadfordICML2023}, which measures speech intelligibility, and
(4) \textit{speaker encoder cosine similarity (SECS)$\uparrow$} by Resemblyzer,\footnote{\url{https://github.com/resemble-ai/Resemblyzer}} which measures speaker similarity.

\subsection{Model analysis}
\label{subsec:model_analysis}

\customparagraph{Component analysis.}
First, we examined the contribution of each component described in Section~\ref{sec:proposal}.
We used \textit{FastVoiceGrad + $\bm{p}_{\phi}$}, where the frozen BNE was simply replaced with a trainable $\bm{p}_{\phi}$, as the baseline, and each component was gradually added to it.
The results in Table~\ref{tab:component_analysis} demonstrate that each component improves performance in most cases.
Two exceptions are as follows:
(1) \textit{FastVoiceGrad + $\bm{p}_{\phi}$} achieved the best CER at the expense of a lower speaker conversion (SECS), and
(2) \textit{Reconversion} degraded SECS compared to \textit{Conversion} owing to the trade-off between content preservation and speaker conversion.
\textit{Inverse} mitigates this trade-off and outperforms both.

\begin{table}[h]
  \vspace{-2mm}
  \caption{Results of component analysis.}
  \vspace{-2mm}
  \label{tab:component_analysis}
  \newcommand{\spm}[1]{{\tiny$\pm$#1}}
  \setlength{\tabcolsep}{3pt}
  \centering
  \scriptsize{
    \begin{tabularx}{\columnwidth}{lCCCC}
      \toprule
      \multicolumn{1}{c}{\texttt{Model}}
      & \texttt{UTMOS$\uparrow$} & \texttt{DNSMOS$\uparrow$} & \texttt{CER$\downarrow$} & \texttt{SECS$\uparrow$}
      \\ \midrule
      FastVoiceGrad + $\bm{p}_{\phi}$
      & 3.45 & 3.64 & \first{0.4} & 0.718
      \\
      + Conversion (Sec.~\ref{subsec:extension})
      & \second{3.98} & 3.76 & 1.4 & \second{0.847}
      \\
      + Reconversion (Sec.~\ref{subsec:reconversion})
      & \first{4.03} & \second{3.79} & 1.3 & 0.844
      \\
      + Inverse (Sec.~\ref{subsec:inverse})
      & \first{4.03} & \first{3.82} & \second{1.2} & \first{0.853}
      \\ \bottomrule
    \end{tabularx}
  }
  \vspace{-2mm}
\end{table}

\smallskip
\customparagraph{Comparison with direct distillation.}
We introduce \textit{ADCD} to prevent $\bm{p}_{\phi}$ from learning simple identity transformations.
As discussed in Section~\ref{sec:introduction}, an alternative solution is \textit{direct distillation}, which trains $\bm{p}_{\phi}$ such that its output matches that of the original content encoder with FastVoiceGrad training.
Table~\ref{tab:comparison_distillation} presents a comparison between these methods when the model capacity of $\bm{p}_{\phi}$ was varied by varying the number of layers.
We found that \textit{FasterVoiceGrad} consistently achieved high scores even with one-layer $\bm{p}_{\phi}$, whereas \textit{direct distillation} with one-layer $\bm{p}_{\phi}$ significantly worsened the scores other than CER.
These results demonstrate the superiority of \textit{FasterVoiceGrad} over \textit{direct distillation}.

\begin{table}[h]
  \vspace{-2mm}
  \caption{Comparison among different distillation methods.}
  \vspace{-2mm}
  \label{tab:comparison_distillation}
  \newcommand{\spm}[1]{{\tiny$\pm$#1}}
  \setlength{\tabcolsep}{1pt}
  \centering
  \scriptsize{
    \begin{tabularx}{\columnwidth}{ccCCCC}
      \toprule
      \multicolumn{1}{c}{\texttt{Model}} & \texttt{\# Layers}
      & \texttt{UTMOS$\uparrow$} & \texttt{DNSMOS$\uparrow$} & \texttt{CER$\downarrow$} & \texttt{SECS$\uparrow$}
      \\ \midrule
      Direct distillation & 1
      & 3.35 & 3.66 & \first{1.2} & 0.767
      \\
      \textit{FasterVoiceGrad} & 1
      & \first{4.01} & \first{3.81} & 1.6 & \first{0.852}
      \\ \midrule
      Direct distillation & 3
      & 3.91 & 3.74 & 2.1 & 0.845
      \\
      \textit{FasterVoiceGrad} & 3
      & \first{4.03} & \first{3.82} & \first{1.2} & \first{0.853}
      \\ \midrule
      Direct distillation & 6
      & 3.92 & 3.75 & 2.0 & 0.846
      \\
      \textit{FasterVoiceGrad} & 6
      & \first{4.04} & \first{3.82} & \first{1.2} & \first{0.854}
      \\ \bottomrule
    \end{tabularx}
  }
  \vspace{-2mm}
\end{table}

\subsection{Comparison with FastVoiceGrad}
\label{subsec:comparison_with_fastvoicegrad}

To assess the validity of \textit{FasterVoiceGrad} against the state-of-the-art one-step diffusion-based VC, we compared it with \textit{FastVoiceGrad}~\cite{TKanekoIS2024}.
For a thorough evaluation, we measured the conversion speed using the real time factor (\textit{RTF}).
Specifically, the processing time of $\bm{\mu}_{\phi}$ and the content encoder was divided by the audio playback time to compute \textit{RTF$_{\text{GPU}}$$\downarrow$} on a single GeForce RTX 4090 GPU and \textit{RTF$_{\text{CPU}}$$\downarrow$} on a single-thread Intel Core i9-14900KF CPU.
We also conducted MOS tests for 90 different speaker/sentence pairs to evaluate speech quality (\textit{qMOS}; five-point scale: 1 = bad, 2 = poor, 3 = fair, 4 = good, and 5 = excellent) and speaker similarity (\textit{sMOS}; four-point scale: 1 = different (sure), 2 = different (not sure), 3 = same (not sure), and 4 = same (sure)).
More than 1,000 responses were collected from 11 participants in each test.
The ground-truth speech and speech converted using DiffVC-30~\cite{VPopovICLR2022} (a commonly used baseline) were included as anchor samples.

The objective scores in Table~\ref{tab:comparison_objective_score} show that \textit{FasterVoiceGrad} performed better than or similarly to \textit{FastVoiceGrad} in all metrics while reducing RTF$_{\text{GPU}}$ and RTF$_{\text{CPU}}$ by factors of 6.6 and 1.8, respectively.
The subjective scores in Table~\ref{tab:comparison_subjective_score} indicate that \textit{FasterVoiceGrad} outperformed \textit{FastVoiceGrad} in terms of speech quality but underperformed in terms of speaker similarity.
A potential reason for the higher SECS but lower sMOS is that \textit{FasterVoiceGrad} achieves sufficient speaker conversion for speaker encoders (both for SECS and $\bm{\mu}_{\phi}$) but falls short perceptually.
Unlike BNE, $\bm{p}_{\phi}$ does not explicitly separate the speaker's identity from the input source mel-spectrogram, causing some remnants of the source speech to persist.
This may be perceptible to humans but not detectable by neural speaker encoders.
Addressing this issue using an improved speaker encoder is a promising direction for future research.

\begin{table}[h]
  \vspace{-1mm}
  \caption{Comparison of objective scores.}
  \vspace{-2mm}
  \label{tab:comparison_objective_score}
  \newcommand{\spm}[1]{{\tiny$\pm$#1}}
  \setlength{\tabcolsep}{0pt}
  \centering
  \scriptsize{
    \begin{tabularx}{\columnwidth}{cCCCCCC}
      \toprule
      \texttt{Model}
      & \texttt{UTMOS$\uparrow$} & \texttt{DNSMOS$\uparrow$} & \texttt{CER$\downarrow$} & \texttt{SECS$\uparrow$} & \texttt{RTF$_{\text{GPU}}$$\downarrow$} & \texttt{RTF$_{\text{CPU}}$$\downarrow$}
      \\ \midrule
      Ground truth
      & 4.14 & 3.75 & 0.1 & 0.871 & -- & --
      \\ \midrule
      FastVoiceGrad
      & 3.96 & 3.77 & 1.3 & 0.847 & 0.00560 & 0.1347
      \\
      \textit{FasterVoiceGrad}
      & \first{4.03} & \first{3.82} & \first{1.2} & \first{0.853} & \first{0.00085} & \first{0.0747}
      \\ \bottomrule
    \end{tabularx}
  }
  \vspace{-2mm}
\end{table}

\begin{table}[h]
  \vspace{-2mm}
  \caption{Comparison of MOS with 95\% confidence intervals.}
  \vspace{-2mm}
  \label{tab:comparison_subjective_score}
  \newcommand{\spm}[1]{{\tiny$\pm$#1}}
  \setlength{\tabcolsep}{0pt}
  \centering
  \scriptsize{
    \begin{tabularx}{0.6\columnwidth}{cCC}
      \toprule
      \texttt{Model}
      & \texttt{qMOS$\uparrow$} & \texttt{sMOS$\uparrow$}
      \\ \midrule
      Ground truth
      & 4.43\spm{0.07} & 3.59\spm{0.08}
      \\
      DiffVC-30
      & 3.58\spm{0.10} & 2.27\spm{0.11}
      \\ \midrule
      FastVoiceGrad
      & \second{3.60\spm{0.09}} & \first{2.84}\spm{0.11}
      \\
      \textit{FasterVoiceGrad}
      & \first{3.81}\spm{0.09} & \second{2.66\spm{0.12}}
      \\ \bottomrule
    \end{tabularx}
  }
  \vspace{-2mm}
\end{table}

\subsection{Versatility analysis}
\label{subsec:versatility_analysis}

To verify the independence of the dataset, experiments were conducted on the LibriTTS dataset~\cite{HZenIS2019}.
The results in Table~\ref{tab:results_libritts} demonstrate a tendency similar to that on the VCTK dataset; that is, \textit{FasterVoiceGrad} achieved objective scores comparable to those of \textit{FastVoiceGrad} while reducing RTF$_{\text{GPU}}$ and RTF$_{\text{CPU}}$ by factors of 6.9 and 1.8, respectively.

\begin{table}[h]
  \vspace{-2mm}
  \caption{Results on LibriTTS dataset.
    $^\dag$Ground-truth converted speech does not necessarily exist in LibriTTS.
    Therefore, source speech was used to calculate the scores.}
  \vspace{-2mm}
  \label{tab:results_libritts}
  \newcommand{\spm}[1]{{\tiny$\pm$#1}}
  \setlength{\tabcolsep}{0pt}
  \centering
  \scriptsize{
    \begin{tabularx}{\columnwidth}{cCCCCCCC}
      \toprule
      \texttt{Model}
      & \texttt{UTMOS$\uparrow$} & \texttt{DNSMOS$\uparrow$} & \texttt{CER$\downarrow$} & \texttt{SECS$\uparrow$} & \texttt{RTF$_{\text{GPU}}$$\downarrow$} & \texttt{RTF$_{\text{CPU}}$$\downarrow$}
      \\ \midrule
      Ground truth$^\dag$
      & 4.06 & 3.70 & 0.6 & -- & -- & --
      \\ \midrule
      FastVoiceGrad
      & 3.94 & 3.75 & \first{1.2} & \first{0.843} & 0.00691 & 0.1424
      \\
      \textit{FasterVoiceGrad}
      & \first{4.02} & \first{3.77} & 1.3 & \first{0.843} & \first{0.00102} & \first{0.0782}
      \\ \bottomrule
    \end{tabularx}
  }
  \vspace{-2mm}
\end{table}

\section{Conclusion}
\label{sec:conclusion}

We proposed \textit{FasterVoiceGrad}, a faster one-step diffusion-based VC model obtained by simultaneously distilling a reverse diffusion model and content encoder with \textit{adversarial diffusion conversion distillation}.
Experimental results demonstrated the importance of each component and the validity against other methods.
Future studies will include applications to advanced or practical VC, such as accent conversion and real-time VC.

\bibliographystyle{IEEEtran}
\bibliography{refs}

\end{document}